\documentclass[prd,preprintnumbers,showpacs,12pt]{revtex4}
\usepackage[active]{srcltx}
\usepackage{epsfig}
\usepackage{graphicx}
\usepackage{bm}
\usepackage{amsmath}
\usepackage{amssymb}
\usepackage{color}

\newcommand{\be}{\begin{equation}}

\newcommand{\ee}{\end{equation}}
\newcommand{\bea}{\begin{eqnarray}}
\newcommand{\eea}{\end{eqnarray}}

\def\f{\frac}

\def\HH{{\cal H}}
\def\TT{{\cal T}}

\newcommand{\nn}{\nonumber}
\newcommand{\de}{\partial}

 \def\slash#1{\setbox0=\hbox{$#1$}#1\hskip-\wd0\dimen0=5pt\advance
       \dimen0 by-\ht0\advance\dimen0 by\dp0\lower0.5\dimen0\hbox
         to\wd0{\hss\sl/\/\hss}}
\usepackage[latin1]{inputenc}

\begin{document}

\title{Non equivalence of Carroll limits in relativistic string theory\\}

\author{Roberto Casalbuoni}\email{casalbuoni@fi.infn.it}
\author{Daniele Dominici}\email{dominici@fi.infn.it}
\affiliation{Department of Physics and Astronomy, University of Florence and
INFN, Via Sansone 1, 50019, Sesto Fiorentino (FI), Italy}
\author{Joaquim Gomis} \email{joaquim.gomis@ub.edu}
\affiliation{
 Departament de F\'isica Qu\`antica i Astrof\'isica 
and Institut de Ci\`encies del Cosmos (ICCUB), Universitat de Barcelona,  Martí i Franquès 1, 08028 Barcelona,
Spain}

\begin{abstract} 
We construct new Carroll strings in flat space by considering the Carroll limit of equivalent relativistic string theories at classical level. In the limit these Carroll strings are no longer equivalent and have different degrees of freedom. This fact is due to a general phenomenon that the configuration space and canonical  Lagrangians are no longer equivalent after a singular redefinition of the variables and of the parameters of the starting Lagrangians.  

\end{abstract}
\pacs{02.20.-a; 02.20.Tw; 03.30.+p}

\maketitle
\section{Introduction}
Carroll symmetry, which was introduced by \cite{Levy-Leblond:1965,gupta:1966} as the limit  of velocity of light going to zero, $c\to 0$, of
Poincaré symmetry,
 has in the recent years received a lot of attention because it arises in different physical systems.
 Carroll structures emerge  when  one considers asymptotically flat spacetimes in gravity \cite{Barnich:2009se,Bagchi:2010zz,Bagchi:2012cy}, where the corresponding symmetry is the  BMS group \cite{Bondi:1962px,Sachs:1962zza} which is isomorphic to the conformal Carroll group \cite{Duval:2014uva}, in 
 the geometry of the black-hole near horizon \cite{Penna:2018gfx,Donnay:2019jiz,Bagchi:2023cfp}, null surfaces \cite{Duval:2014uoa}, relativistic fluids \cite{Ciambelli:2018wre}  and in condensed matter \cite{Bidussi:2021nmp,Marsot:2022imf}. Very recently Carrollian amplitudes have been computed from strings \cite{Stieberger:2024shv}.
 An introductory review of some  aspects of non-Lorentzian 
theories can be found in \cite{Bergshoeff:2022eog}.
Dynamical realizations of the
conformal Carroll structures \cite{Duval:2014uva} have been studied, for example, through   the construction of  Carrollian particles. A massive non-conformal Carroll particle
was introduced  as Carroll limit of relativistic massive particle
\cite{Bergshoeff:2014jla} and  can also be obtained by the technique  of the  coadjoint orbits of the Carroll group \cite{Duval:2014uoa}. The massless Carroll particle was also considered in \cite{Bergshoeff:2014jla}.

 The symmetries of massive and massless  free Carroll particles are infinite dimensional. In the massless case the symmetry contains the finite conformal Carroll group introduced in \cite{Bagchi:2016bcd,Chen:2021xkw}  and also the BMS symmetry \cite{Bondi:1962px,Sachs:1962zza}.
 In the case of two non-conformal  interacting particles the symmetry algebra is finite dimensional \cite{Bergshoeff:2014jla}.
 Recently two models of two interacting 
conformal Carroll particles \cite{Casalbuoni:2023bbh}, which can be obtained as the 
Carrollian limit of two relativistic conformal particles \cite{Casalbuoni:2014ofa}, have been 
 constructed. The first model describes particles  that do not move and exhibits 
infinite dimensional symmetries which are  reminiscent of the BMS symmetries.
 A second model was also  proposed, where
the particles have non zero velocity 
 and therefore, as a consequence of the limit $c\to 0$,  are tachyons.  Infinite dimensional symmetries are also present  in this model.


Some preliminary work on Carroll strings have been already performed in \cite{Cardona:2016ytk} by considering the formulation in the phase space configuration of the Nambu Goto string and by \cite{Bagchi:2023cfp} where the near horizon property of the string are investigated.
We extend the work contained in   \cite{Cardona:2016ytk} by studying alternative but equivalent formulations of the relativistic string
and show that the  particle Carroll limit  gives rise to Carroll theories with  different number of degrees of freedom but with  the same  
 Carroll algebra.  Analogous considerations can be done for the Carroll string limit \cite{Cardona:2016ytk}.

We also consider the {\it tachyonic} relativistic strings   
which, in the Carrol limit, produce also new Carroll strings in flat space time.
As a warmup exercise, in appendix we recall the Carroll limit of the different formulations  of the relativistic massive particle, including the tachyon case.

The organization of the paper is as follows. In section II we discuss the non equivalence of the configuration and canonical Lagrangians in a singular limit, in section III we analyze different formulations of the bosonic relativistic string and their Carroll  particle limits and in appendix we show the non equivalence for the Carroll limits of different formulations of the relativistic particle.

\section{Non equivalence of the configuration space and canonical Lagrangians in a singular limit}

Since in this paper we are interested in discussing the non equivalence of Carroll limits, we start by considering a general
dynamical system described by a set of variables discrete or continuous and by a Lagrangian depending on these variables and on their time derivatives. In order to fix the ideas we will consider discrete variables but the  considerations that we will do can be easily extended  to the continuum case.
The equations of motion for such a system can be obtained by looking for the extrema of the action functional
\be
S=\int dt\, L(q_i,\dot q_i,t),~~~i=1\cdots N.  \label{eq:1}
\ee
Such a system can also be described  by introducing the Hamiltonian through a Legendre transformation and eliminating the velocities in terms of the momenta. This inversion is not always possible 
 in the full phase space
leading to  constraints that determine a submanifold where this inversion is possible. In this instance we suppose to follow the Dirac procedure solving the second class constraints (or using Dirac brackets) and adding to the canonical Hamiltonian the first class constraints times arbitrary functions \cite{DiracP.A.M1964Loqm}. In any case there exists a  well defined procedure to get the Hamiltonian from the Lagrangian \footnote{The equivalence among the Lagrangian and Hamiltonian formalism has been studied. See for example \cite{Batlle:1985ss}.}, and this is the point  we will be interested in. Therefore, in the following general discussion, for sake of simplicity, we will suppose that there are no constraints. 

In the Hamiltonian formalism  one requires the equations of motion that are equivalent to the Lagrange equations coming from the variation of Eq. (\ref {eq:1}).
As it is well known, it is possible to derive the Hamilton equations from a variational principle, introducing an extended Lagrangian and a corresponding action defined in a space spanned by $q_i, \dot q_i,p_i$:
\be
S_E=\int dt\, L_E(q_i,\dot q_i, p_i,t)= \int dt\, \left(\sum_{i=1}^N p_i \dot q_i- H(q_i, p_i,t)\right)_
,~~~i=1\cdots N.  \label{eq:2}
\ee
We would like to emphasise the fact that the two formulations corresponding to Eqs. (\ref{eq:1}) and 
(\ref{eq:2})  are completely equivalent.
The equations of motion of both Lagrangians
coincide when we use the Legendre transform.

Due to this equivalence, if we perform some transformation of the variables in $L$, this will induce a corresponding variation in $L_E$ such that the two will remain equivalent. However the situation changes completely if our transformation is singular. 
Just to  exemplify, suppose that we scale a single variable $q_i\to {q_i}/\omega$  and suppose that $L$ is finite, maybe by some $\omega$ redefinition of certain parameters appearing in the Lagrangian, for $\omega\to\infty$. Then, when we evaluate $L_E(\omega)$, starting from $L(\omega)$ and send $\omega\to \infty$, in general we have to make some further $\omega$-dependent transformation on $L_E(\omega)$ to make it finite and in general inequivalent:
\be
\lim_{\omega\to\infty}L(\omega) \not\approx \lim_{\omega\to\infty}\tilde L(\omega),
\ee
where $\tilde L(\omega)$ is $L_E$ transformed in such way to get a finite result.  Note that in some cases to get a finite limit it is also necessary to add some additional terms to the Lagrangian  
\cite{Gomis:2000bd,
Gomis:2005pg}.

A further observation concerns the case in which the Lagrangian $L$ is invariant under a continuous group of transformations, $G$, then, as before, we will rescale one or more variables in such a way that
$
\lim_{\omega\to\infty}L(\omega)
$
is finite. Then, in general, the resulting Lagrangian is invariant under a contraction of the Lie algebra of the original symmetry. Of course, the Lagrangian $L_E$ has the same group of symmetries $G$ as $L$ and the question arises about the symmetries of  $\lim_{\omega\to\infty}\tilde L(\omega)$. Generally speaking 
the two limiting cases are invariant under the same contracted symmetry group, but corresponding to two different realisations. 

\section{Carroll strings from relativistic strings}
\subsection{Canonical action from a tension full string }
Let us consider the relativistic Nambu Goto string in the phase space formulation; the action is given 
by
\be
S=\int d \tau\int_0^{2\pi} d\sigma {\cal L}_E
\ee
where the Lagrangian density is given by
\be\label{relcanonical}
{\cal L}_E=p\cdot x- \f e 2 (p^2+T^2 x^{\prime 2})-\mu \,p \cdot x^\prime,
\ee
where $p^\mu\equiv p^\mu(\tau, \sigma)$, $x^\mu\equiv x^\mu(\tau, \sigma)$, $ T$ is the string tension and $e(\tau,\sigma),\mu(\tau,\sigma)$ are Lagrange multipliers. 

 In this case, to perform the particle Carroll limit,  we introduce the invertible change 
\be\label{limitcarroll}
x^0= \f t \omega,\quad p^0=\omega E, \quad e=\f e {\omega^2},\quad T=\omega  T
\ee
and we consider $\omega\to\infty$, as already done in \cite{Cardona:2016ytk},  and we obtain
\be
\lim_{\omega\to\infty} {\tilde {\cal L}}_E(\omega)=-E\dot t+ \bm p\cdot \dot {\bm x}-\f e 2(-E^2+ T^2 {\bm x}^{\prime 2})-\mu (-Et^\prime +{\bm p}\cdot {\bm x}^\prime). \label{eq:39}
\ee
The corresponding first class constraints are
\be
\Pi_e=\f {\de {\cal L}}{\de \dot {e}}=0,\quad \Pi_\mu=\f {\de{\cal L}}{\de \dot {\mu}}=0,\quad {\rm primary},
\ee
\be
-E^2+ T^2 {\bm x}^{\prime 2}=0,\quad Et^\prime -{\bm p}\cdot {\bm x}^\prime=0,\quad {\rm secondary}.
 \label{eq:41}
\ee
The Lagrangian and the constraints, Eq. (\ref{eq:39}) and Eqs. (\ref{eq:41}), are invariant under the Carroll transformations 
\bea
&&\delta t={\bm\beta}\cdot {\bf x}+a_t\,,\hskip 2truecm
\delta x^i= \epsilon^{ijk}\theta^j x^k+a^i\,,
\nn\\
&& \delta p^i= \epsilon^{ijk}\theta^j p^k +{\beta^i} E\,, \hskip 2.9truecm
\delta E=0\,,
\label{eq:10}\eea
 and assuming $\delta_C e=\delta_C \mu=0$.
 The  Lagrangian (\ref{eq:39}) has an infinite set of symmetries that contain Carroll  \cite{Cardona:2016ytk}.

As already shown in \cite{Cardona:2016ytk}
 by going in the conformal gauge $e=1,\mu=0$,
  the dynamics of this Carroll string is trivial, since the spatial coordinates of the string are constant in $\tau$. However the momentum density is not constant.

The number of degrees of freedom is given by  $N_{dof}=1/2 [ 2(D+2)-2\times 4]=D-2$ which corresponds to the transverse degrees of freedom
of  string.

Notice that the Lagrangian eq. (\ref{eq:39}) is related to the Lagrangian of the non-vibrating non-relativistic string by the the mapping \cite{Batlle:2016iel}
\be
E\to {\bm P},\quad  {\bm P}\to -E ,\quad {\bm x}\to - t.
\ee

\subsection{Lagrangian in configuration space and with einbeins}
\label{subsb}

If we eliminate the momenta of the relativistic string Lagrangian (\ref{relcanonical}) we obtain the corresponding Lagrangian in the configuration space formulated using einbeins variables
\be
{\cal L}=\f  1 {2e} \dot x^2-\f \mu e \dot x \cdot x^\prime+  \f 1 2 \f {\mu^2}{e} x^{\prime 2}- \f {T^2}2 e x^{\prime 2}.
\label{ltachion}
\ee
In this case, the Carroll limit is
\be\label{carrolltras0}
x^0= \f t \omega, \omega\to\infty.
\ee
We obtain
\be\lim_{\omega\to\infty} { {\cal L}}(\omega)=\f  1 {2e} \dot {\bm x}^2-\f \mu e \dot {\bm x }\cdot {\bm x}^\prime +  \f 1 2 \f {\mu^2}{e} {\bm x}^{\prime 2}- \f {T^2}2 e {\bm x}^{\prime 2}.
\ee
Note that the tension is not rescaled.

We have
\be
\Pi_e=0, 
\Pi_\mu=0, 
{E}=\f {\de {\cal L}}{\de \dot {t}}=0,{\rm primary\, constraints},
\label{eq:15}
\ee

\be
{\bm p}=\f {\de {\cal L}}{\de \dot {\bm x}}=\f 1 e (\dot {\bm x}-\mu{\bm x}^\prime),
\ee
\be
0=\f {\de {\cal L}}{\de e}=-\f 1 2 \left [ (\dot {\bm x}-\mu {\bm x}^\prime)^2 +T^2  {\bm x}^{\prime 2}\right ],
\ee
\be
0=\f {\de {\cal L}}{\de \mu}=-e (\dot {\bm x}-\mu {\bm x}^\prime)\cdot  {\bm x}^\prime.
\ee
The canonical Hamiltonian is
\be
H_c=\f e 2({\bm p}^2+T^2 {\bm x^{\prime 2}})+\mu 
 \,{\bm p}\cdot {\bm x}^\prime.
\ee

The secondary constraints are
\be
{\cal H}={\bm p}^2+T^2 {\bm x^{\prime 2}}=0,\quad {\cal T}={\bm p}\cdot {\bm x}^\prime =0.
\label{eq:53a}
\ee
Assuming $T\neq 0$, the first constraint of Eq.~(\ref{eq:53a}) implies
\be
{\bm p}^2={\bm x^{\prime 2}}=0.
\label{irreg}
\ee
In conclusion we have six first  class constraints; therefore the number of degrees of freedom of the system is $N_{dof}=1/2 [ 2(D+2)-2\times 6]=D-4$.

Notice that the constraints (\ref{irreg}) are irregular ones \cite{Miskovic:2003ex}. If we substitute them 
 by
the linearized constraints
\be
{\bm p}={\bm x}^\prime=0,
\label{irreg1}
\ee
the  constraints (\ref{irreg1}) become now second class. Therefore taking into account the three first class constraints of Eq.~(\ref{eq:15})
and the $D-1$ second class ones of (\ref{irreg1}), the number of degrees of freedom becomes $N_{dof}=1/2 [ 2(D+2)-2\times 3-2\times (D-1)]=0$. Proceeding in this way no  local degree of freedom is left.

The same result is obtained by performing the 
 Carroll particle limit eq. (\ref{carrolltras0})
limit in the Nambu Goto Lagrangian
\be
{\cal L}=-T\sqrt{(\dot x\cdot x^\prime)^2-{\dot x}^2 x^{\prime 2}},
\ee
which gives
\be
{\cal L}=-T\sqrt{({\dot {\bm x}}\cdot {\bm x}^\prime)^2-\dot {\bm x}^2 \bm x^{\prime 2}},
\ee
from which the constraints (\ref{eq:53a}) can be directly obtained. 
The Carroll transformations  are given by eqs. (\ref{eq:10})
for the space time variables.

\subsection{Canonical Lagrangian from a tachyonic string}

To perform the Carroll limit of the tachyonic string 
\footnote{The non-relativistic limit of a tachyonic relativistic particle in  canonical form  was done in \cite{Batlle:2017cfa}.}
in the phase space formulation it is necessary to change $T^2\to -T^2$  
\be\label{relcanonicaltach}
{\cal L}_E=p\cdot x- \f e 2 (p^2-T^2 x^{\prime 2})-\mu \, p \cdot x^\prime.
\ee
The Carroll limit (\ref{limitcarroll}) gives
\be
\lim_{\omega\to\infty} {\tilde {\cal L}}_E(\omega)=-E\dot t+ \bm p\cdot \dot {\bm x}-\f e 2(-E^2- T^2 {\bm x}^{\prime 2})-\mu (-Et^\prime +{\bm p}\cdot {\bm x}^\prime). \label{eq:43}
\ee
As it is clear from the form of the quadratic constraint $-E^2- T^2 {\bm x}^{\prime 2}=0$, if we assume $T\neq 0$  that we have the constraints $E^2=
 {\bm x}^{\prime 2}=0$.
  
  Since $\Pi_e=0$, $\Pi_\mu=0$, $E=0$, $\bm x^{\prime 2}=0$, ${\bm p}\cdot {\bm x}^\prime=0$, 
   the number of degrees of freedom is
  $N_{dof}=1/2 [2(D+2)-2\times 5]=2(D-3)$.
 
 If we consider the linear constraint ${\bm x}^{\prime}=0$ associated to the irregular constraint ${\bm x}^{\prime 2}=0$
\cite{Miskovic:2003ex}, the  counting of degrees of freedom goes as $N_{dof}=1/2 [2(D+2)-2(3+D-1)]=0$. 
Therefore, as in a previous case, there are no local degrees of freedom.

\subsection{Tachyonic string Lagrangian in the configuration space}
The tachyonic string is obtained from (\ref{ltachion})
again if we change $T^2\to -T^2$ obtaining

\be
{\cal L}=\f  1 {2e} \dot x^2-\f \mu e \dot x\cdot  x^\prime+  \f 1 2 \f {\mu^2}{e} x^{\prime 2}+ \f {T^2}2 e x^{\prime 2}.
\ee
By performing the Carroll limit (\ref{carrolltras0})
we obtain
\be\lim_{\omega\to\infty} { {\cal L}}(\omega)=\f  1 {2e} \dot {\bm x}^2-\f \mu e \dot {\bm x }\cdot {\bm x}^\prime+  \f 1 2 \f {\mu^2}{e} {\bm x}^{\prime 2}+ \f {T^2}2 e {\bm x}^{\prime 2}.
\ee
Proceeding as in the case of subsection \ref{subsb} we obtain
as  secondary constraints 
\be
{\cal H}={\bm p}^2-T^2 {\bm x^{\prime 2}}=0,\quad {\cal T}={\bm p}\cdot {\bm x}^\prime =0,
\label{eq:53}
\ee
which are first class.

Their algebra is given by
\be
\{ \HH(\sigma),\HH(\sigma')\}=4T^2\left [\TT(\sigma')\de_{\sigma'}\delta(\sigma-\sigma')-\TT(\sigma)\de_{\sigma}\delta(\sigma-\sigma')\right ],
\ee
\be
\{ \HH(\sigma),\TT(\sigma')\}=\HH(\sigma)\de_{\sigma}\delta(\sigma-\sigma'),
\ee
\be
\{ \TT(\sigma),\TT(\sigma')\}=4T^2\left [-\TT(\sigma')\de_{\sigma'}\delta(\sigma-\sigma')+\TT(\sigma)\de_{\sigma}\delta(\sigma-\sigma')\right ].
\ee
 We have three primary first class
$\Pi_e=0, \Pi_\mu=0, E=0$, and two secondary first class constraints, therefore we have three 
type gauge transformations. We have an arbitrary transformation in the time coordinate and two dimensional euclidean isomorphism.
 Taking into account the presence of five first class constraints, the number of degrees of freedom is given by $N_{dof}=1/2 [2(N+2)-2\times 5]=D-3$.

We can check this statement explicitly if we introduce a gauge fixing $t=0$ to make the first class constraint $E=0$ second class; then  the reduced
phase space is (${\bm x}, {\bm p}$) and  the canonical Lagrangian is given by
\begin{equation}\label{psact1}
{\cal L}=   \dot{\bf x}\cdot{\bf p} - \frac{1}{2} e\left ({\bf p}^2 - T^2 {\bf x}^{\prime 2}\right )- \mu {\bf p}\cdot {\bf x}'\, .
\end{equation}
Note the difference with respect to the non-vibrating non-relativistic string \cite{Batlle:2016iel}.

The action is still invariant under reparametrizations of worldsheet $\tau$ and $\sigma$  coordinates. At least locally,  this allows us to identify  one of the space 
coordinates with $\tau$ and we may then choose the unit of length such that 
\begin{equation}\label{gfix}
x^1=\tau,\quad x^2=\sigma
\end{equation}
and assume a $2\pi$ periodicity for the variable $x^2$
\be
x^2\equiv x^2+2\pi.
\ee
With these choices the constraints (\ref{eq:53}) become second class. Furthermore we have 
\be
{\cal L}=   \dot{\bf x}\cdot{\bf p} =p^1+\sum_{a=3}^{D-1}{\dot x}^a p^a.
\label{eq:37}
\ee
The constraints (\ref{eq:53}) become equivalent to 
\be
p^2+\sum_{a=3}^{D-1}p^ax^{a\prime}=0,
\ee
\be
{\bm p}^2 -T^2 \left [1+\sum_{a=3}^{D-1}(x^{a\prime})^2\right ]=0,
\ee
or
\be
p^2=-\sum_{a=3}^{D-1}p^ax^{a\prime},
\ee
\be
p^1=\pm T\sqrt{1+\sum_{a=3}^{D-1}(x^{a\prime})^2-T^{-2}\sum_{a=3}^{D-1}[(p^ax^{a\prime})^2+  (p^{a})^2]}\,.
\ee
By choosing the minus sign solution, from eq.~(\ref{eq:37}) we can extract  the Hamiltonian density
\be
H=T\sqrt{1+\sum_{a=3}^{D-1}(x^{a\prime})^2-T^{-2}\sum_{a=3}^{D-1}\left [(p^ax^{a\prime})^2+  (p^{a})^2\right ]}\,.\ee

This string is propagating in euclidean space,  we could call it an instantonic string. By double dimensional reduction we recover the tachyonic Carroll particle with zero energy \cite{deBoer:2021jej}.

\section{Conclusions}
We have constructed new Carroll strings in flat space by considering the  Carroll particle limit of equivalent relativistic string theories at classical level. In the limit these Carroll strings are no longer equivalent and have different degrees of freedom.
This non equivalence should be a general property
for other limits \cite{Brugues:2004an,Brugues:2006yd,
Barducci:2018wuj,Casalbuoni:2023bbh} of free and interacting particles, p-branes and field theories, in particular about relativistic gravity. 

The relation with other approaches to construct Carroll theories like the seed procedure \cite{Bergshoeff:2022qkx} should be analyzed.

 \appendix
 
 \section{Relativistic particles}\label{sec:2}

Relativistic time-like particles have tree possible different actions, the configuration action with only spacetime coordinates 
\be\label{particle1}
S=-m\int d\tau\sqrt{-{\dot x}^2},
\ee
where $m$ is the mass of the particle; the metric $g_{\mu\nu}=(-,+,\cdots,+)$ and natural units are assumed.
The action with space-time variables and an einbein variable (Polyakov) formulation is
\be
S=\int d\tau(\frac 1{2e}{\dot x}^2-\frac 12 e\,m^2), \label{eq:1a}
\ee
while the canonical action is
\be\label{particle2}
S=\int d\tau\left [p\cdot\dot x-\frac 1 2 e(p^2+m^2)\right ].
\ee
They are equivalent by elimination of the momenta and the einbein variables of the canonical action. The analogous actions for the 
tachyons are obtained by changing the sign of the square root and changing $m^2\to -m^2$ in the Polyakov and canonical action.
This equivalence is broken when we consider the Carroll  
limit. The causal structure admits two type of time intervals
 that coincide with light-like and space-like intervals.

\subsubsection{Carroll limit  (time-like case)}

Let us consider the Carroll limit of a massive relativistic
particle, whose action is given by Eq.~ (\ref{particle1}), trough the substitution ($c=1$)
\be
x^0\to\frac 1\omega t. 
\label{eq:a14}
\ee
In the limit $\omega\to\infty$ the Lagrangian is not real and we do not proceed furthermore.

Let us now consider Eq. (\ref{eq:1a}) in the limit (\ref{eq:a14}); if we do not scale the einbein and the mass we get
\be
\lim_{\omega\to\infty} L(\omega)= \frac 1{2e}({\dot{\bm x}}^2-e^2m^2).
\label{eq:16}
\ee
There are two primary constraints
\be
E=\f {\de L}{\de \dot t}=0, \Pi_e=\f {\de L}{\de \dot e}=0,
\ee
where $E$ is the energy and $\Pi_e$ is the momentum of the einbein variable $e$. There is also
a secondary constraint
\be
{\bm p}^{\,2}\,+\,m^2=0.
\ee  
which has real solution only when $m=0$.
In this case the 
${\bm p}^{\,2}=0$ constraint is an irregular one \cite{Miskovic:2003ex}.

The associated canonical Lagrangian is 
\be\label{particle5}
L_c=-E\dot t+{\bm p}\cdot\dot{\bm x}-\frac e2
{\bm p^2}-\mu E.
\ee
The Lagrangian coincides with the one of the massless Galilean \cite{souriau} with color equal to zero and zero energy, see \cite{Gomis:2022spp}; the particle does not move and the momentum is equal to zero.
Taking into account the presence of four first class constraints, the number of configuration degrees of freedom  is
  $N_{dof}= 1/2 [2(D+2)-2\times 4]=D-2$. The Lagrangian is invariant under the following boost Carroll transformations
   \bea
&&\delta_C t={\bm\beta}\cdot {\bf x}\,,
\quad\delta_C x^i= 0,
\nn\\
&& \delta_C p^i= {\beta^i} E\,, \quad
\delta_C E=0\,,
\nn\\
 &&\delta_C e=0,\quad  \delta_C \mu=
 - e\, {\bm\beta}\cdot{\bm p}.
 \label{eq:40a}\eea

If we consider the linear constraint ${\bm p}=0$ associated to the irregular constraint ${\bm p^2}=0$,
\cite{Miskovic:2003ex},
the canonical Lagrangian for the linear constraint action becomes
\be\label{particle6}
L_c=-E\dot t+{\bm p}\cdot\dot{\bm x}-{\bm \lambda}\cdot
{\bm p}-\mu E,
\ee
where ${\bm \lambda}$ are $D-1$ Lagrangian multipliers.
There is no a configuration space action.
 The number of degrees of freedom in configuration space is  $N_{dof}= 1/2 [2(D+D-1+1)-2(D-1+D-1+1+1)]=0$.
  In conclusion the model 
 has no local physical degrees of freedom and its
  action is Carroll invariant under the following 
  boost transformations 
  \bea
&&\delta_C t={\bm\beta}\cdot {\bf x}\,,
\quad\delta_C x^i= 0,
\nn\\
&& \delta_C p^i= {\beta^i} E\,, \quad
\delta_C E=0\,,
\nn\\
 &&\delta_C {\bm \lambda}=0,\quad  \delta_C \mu=
 -{\bm\beta}\cdot{\bm\lambda}.
 \label{eq:40}\eea

The Carroll limit of the canonical action,  
Eq. (\ref{particle2}), is defined by
\be
x^0\to\frac 1\omega t, \quad m=\omega M,\quad
e=\frac{\tilde e}{\omega^2}.
\label{eq:14}
\ee
Note the scaling of the einbein variable and of the mass.
The Lagrangian is given by \cite{Bergshoeff:2014jla}
\be
L_c=-E\dot t+{\bm p}\cdot\dot{\bm x}+\frac{\tilde e}2(E^2-\tilde m^2)
\label{eq:24}
\ee
and describes a particle  with constant energy
\be
E^2=\tilde m^2,
\ee
zero velocity and no relation among the momentum  and the velocity.
 The number of degrees of freedom is equal to $N_{dof}=1/2 [ 2(D+1)-2\times 2]=D-1$.
 The Lagrangians (\ref{particle5}), (\ref{particle6})
 and (\ref{eq:24}) are inequivalent.

\subsubsection{Carroll limit  (space-like or tachyon case)}

The configuration relativistic action for a tachyon is 
\be\label{particle3}
S=-m\int d\tau\sqrt{{\dot x}^2}
\ee
and the Carroll limit, Eq (\ref{eq:a14}), gives
\be\label{particle4}
S=-m\int d\tau\sqrt{{\dot {\bm x}}^2}.
\ee
The primary constraints are
\be
{\bm p}^2-m^2=0,\quad E=0.
\ee
Therefore the Lagrangian coincides with the massless Galilean particle with $m$ color and vanishing energy \cite{souriau}.

Let us now consider the relativistic tachyon Lagrangian with einbein variable 
\be
L_T=\frac 1{2e}{\dot x}^2+\frac 12 e\,m^2.
\label{eq:a18}
\ee
By performing the Carroll limit of Eq.~ (\ref{eq:a18}) we obtain
\be
\lim_{\omega\to\infty} L_T(\omega)= \frac 1{2e}({\dot{\bm x}}^2+e^2m^2).
\ee
 There are two primary constraints 
\be
E=0, \Pi_e=0
\ee
and a 
secondary constraint 
\be
{\bm p}^{2}-m^2=0,
\ee
fixing the modulus of the spatial momentum without any restriction of the parameter $m$. Note that the constraints coincide with those of the configuration action and therefore in this case the two limits are equivalent.

This particle is a Carroll tachyon with zero energy and momentum $\bm p$. Also its velocity ${\dot{\bm x}}$ is not zero. In conclusion, taking into account the presence of three first class constraints, the number of  degrees of the physical freedom is equal to $N_{dof}=1/2 [ 2(D+1)-2\times 3]=D-2$.

In analogous way the limit of the tachyon canonical  Lagrangian
\be
 L_{TE}=p\cdot\dot x+\frac 1 2 e(p^2-m^2)
 \ee
 gives
\be
\lim_{\omega\to \infty} \tilde L_{TE}=-E\dot t+{\bm p}\cdot\dot{\bm x}+\frac{\tilde e}2(E^2+\tilde m^2)
\label{eq:30}
\ee
where
\be
\tilde e= e\omega^2,~~~\tilde m^2=\frac{m^2}{\omega^2}.
\ee
We get the constraints
\be
\Pi_{\tilde e}=0,\quad E^2+\tilde m^2=0
\ee
with solutions only for $\tilde m=0$. The particle has zero energy and momentum.
 The number of degrees of freedom is equal to $N_{dof}=1/2 [ 2(D+1)-2\times 2]=D-1$.
We have therefore shown that also in this case there is no equivalence between the formulations which make use of  the configuration space action and  of the canonical one.

\acknowledgments
We would like to thank Diego Hidalgo and Axel Kleinschmidt for useful comments. One of us (JG) would like to thank the Galileo Galilei Institute for Theoretical Physics and the INFN for partial support  where the initial steps on this work was done. The research of JG was supported in part by PID2022-136224NB-C21
and by the State Agency for Research of the Spanish Ministry of Science and Innovation through the Unit of Excellence Maria de Maeztu 2020-2023 award to the Institute of Cosmos Sciences (CEX2019- 000918-M).

--


\end{document}